\documentclass[conference, 10pt]{IEEEtran}
\IEEEoverridecommandlockouts
\usepackage{cite}
\usepackage{amsmath,amssymb,amsfonts}
\usepackage[boxed]{algorithm}
\usepackage[noend]{algpseudocode}
\usepackage{graphicx}
\usepackage{textcomp}
\usepackage{xcolor}
\usepackage{wrapfig}
\usepackage{float}
\usepackage{subcaption}
\usepackage{paralist}
\usepackage{soul}
\def\BibTeX{{\rm B\kern-.05em{\sc i\kern-.025em b}\kern-.08em
		T\kern-.1667em\lower.7ex\hbox{E}\kern-.125emX}}
\begin{document}
	
% 	\title{A Framework for Evaluating Connected Vehicle Security against False Data  Injection Attacks}
		\title{A CAD Framework for Simulation of Network Level Attack on Platoons}
		
	 \author{\IEEEauthorblockN{1\textsuperscript{st} Ipsita Koley}
	 \IEEEauthorblockA{\textit{CSE} \\
	 \textit{Indian Institute of Technology Kharagpur}\\
	ikipsita@gmail.com}
	 \and
	 \IEEEauthorblockN{2\textsuperscript{nd} Sunandan Adhikary}
	 \IEEEauthorblockA{\textit{CSE} \\
	 \textit{Indian Institute of Technology Kharagpur}\\
	 mesunandan@gmail.com}
	 \and
	 \IEEEauthorblockN{3\textsuperscript{rd} Rohit}
	 \IEEEauthorblockA{\textit{CSE} \\
	 \textit{Indian Institute of Technology Kharagpur}\\
	 rohitgns123@gmail.com}
  \and
	 \IEEEauthorblockN{ \textsuperscript{ }}
	 \IEEEauthorblockA{\textit{} \\
	 	\textit{}\\
	 	}
	 \and
	 \IEEEauthorblockN{4\textsuperscript{th} Soumyajit Dey}
	 \IEEEauthorblockA{\textit{CSE} \\
	 \textit{Indian Institute of Technology Kharagpur}\\
	 soumyajit.dey@gmail.com}
	 }
	
	\maketitle
	
	\begin{abstract}
		Recent developments in the smart mobility domain have transformed automobiles into networked transportation agents helping realize new age, large-scale intelligent transportation systems (ITS). The motivation behind such networked transportation is to improve road safety as well as traffic efficiency. In this setup, vehicles can share information about their speed and/or acceleration values among themselves and infrastructures can share traffic signal data with them. This enables the connected vehicles (CVs) to stay informed about their surroundings while moving. However, the inter-vehicle communication channels significantly broaden the attack surface. The inter-vehicle network enables an attacker to remotely launch attacks. An attacker can create collision as well as hamper performance by reducing the traffic efficiency. Thus, security vulnerabilities must be taken into consideration in the early phase of CVs’ development cycle. To the best of our knowledge, there exists no such automated simulation tool using which engineers can verify the performance of CV prototypes in the presence of an attacker. In this work, we present an automated tool flow that facilitates false data injection attack synthesis and simulation on customizable platoon structure and vehicle dynamics. This tool can be used to simulate as well as design and verify control-theoretic light-weight attack detection and mitigation algorithms for CVs. 
	\end{abstract}
	
	
	\section{Introduction}
	\label{secIntroduction}
	In the last few years, we have observed a mammoth development in intelligent transportation systems (ITS). Today's vehicles that participate in building such an advanced ITS, are designed as collection of sophisticated control systems that provide a plethora of features related to safety, performance, power management, comfort, and entertainment. 
	Thus, the ECUs, sensors and the intra-vehicular communication network together form an electrical/electronic (E/E) architecture \cite{moller2019guide} which provides the base for a feature-rich present-day automobile.  
	Though these control programs enhance the performance and safety of an individual automotive but are not enough to ensure efficiency and safety on-road traffic, which is one of the primary goals of ITS. According to data published by the U.S. Department of Transportation’s National Highway Traffic Safety Administration (NHTSA), approximately $31,720$ people died in vehicle crashes from January, $2021$ to September $2021$ \cite{nhsta2021}. These facts motivate the need of connected vehicles (CVs) that can communicate among themselves and coordinate as required in an ITS setup. 
	
	\par 
	% Such advancements in automotives are not limited to features of an individual vehicles. 
	Recent developments in the smart mobility domain have transformed automobiles into  networked transportation agents helping realise new age, large-scale ITS. The motivation behind the networked transportation is to improve the road safety as well as traffic efficiency. In this setup, vehicles can share information about their speed and/or acceleration values among themselves and infrastructures can share traffic signal data with them. This enables the CVs to stay informed about their surroundings while moving. A vehicular ad-hoc network (VANET) \cite{anwer2014survey} is established where these vehicles act as moving nodes/routers to form a mobile network that can help On-Board Units (OBU) in vehicles communicate with other OBUs and road-side units (RSU). In case of standalone vehicles, they share information wirelessly with their surrounding vehicles in order to cooperate each other. Whereas in a connected vehicle platoon, these wireless connections follow certain network topologies that specify which platoon members to share the information with. There are dedicated standards developed for Wireless Access in Vehicular Environments (WAVES). For example, DSRC (Dedicated Short Range Communication) \cite{kenney2011dedicated} enables V2V (Vehicle-to-Vehicle) and V2I  (Vehicle-to-Infrastructure) communications (in general V2X). DSRC is mainly responsible for carrying basic safety messages (BSMs) like velocity and acceleration data from the preceding vehicles (as per the network topology) to enable limited autonomy drive-assist systems like Cooperative Adaptive Cruise Control (CACC). CACC technology is preferred in different ITS applications like vehicle platooning, co-operative merging of vehicles at highway intersections, co-operative driving on signalized corridors, etc. DSRC is used in CACC for its distributed nature and increased reliability. 
	
	\par
\begin{figure}[!h]
		\centering
		\includegraphics[clip,width=0.9\columnwidth]{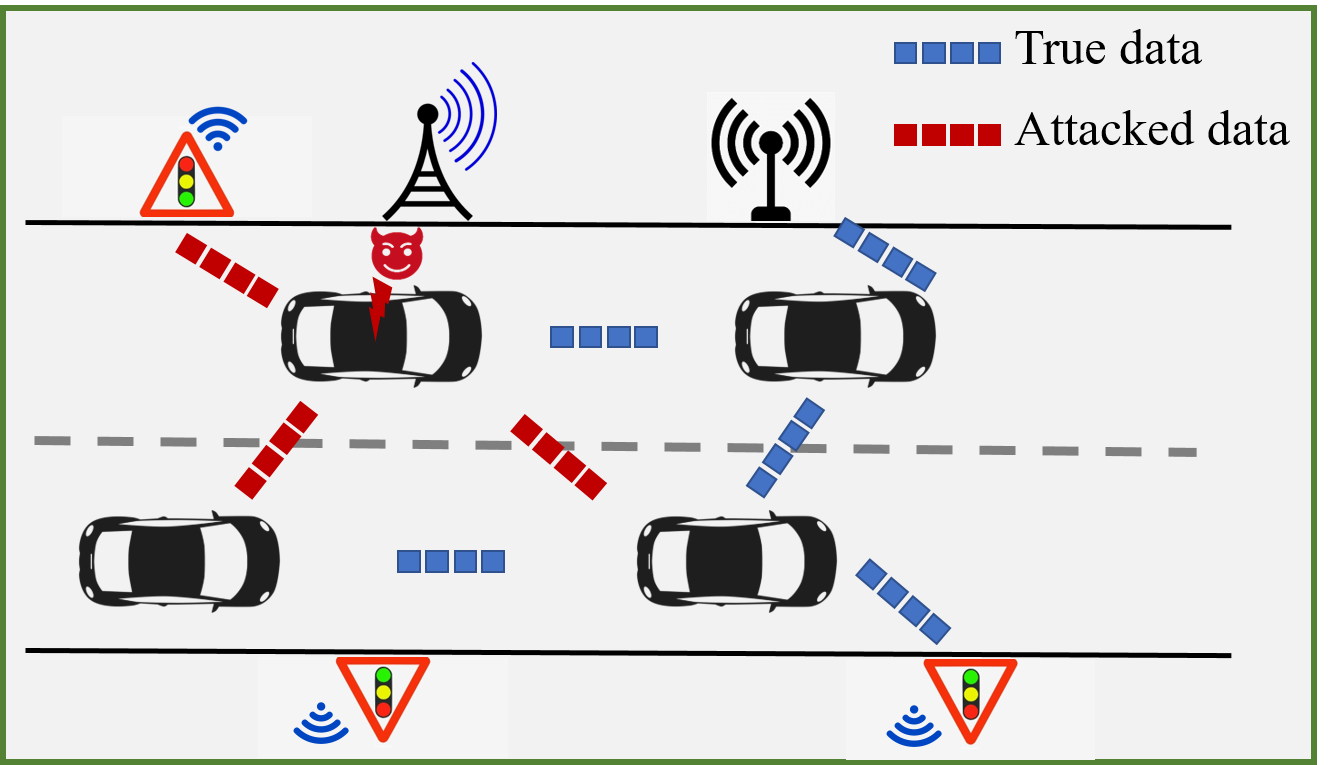}
		\caption{FDI Attack on Connected Vehicles}
		\label{figPlatoonAtk}
	\end{figure}

 Such sophistication in intelligent transportation comes at the cost of security vulnerabilities. Similar to in-vehicle automotive control systems \cite{checkoway2011comprehensive}, CVs are also not free from attacks. The inter-vehicle communication channels significantly expose the intra-vehicular network, thus  broadening  the attack surface. Even without direct physical access to a vehicle's internal network, an attacker can remotely hijack the vehicle's control from the driver.
%	\sa{before going to insider attacks we need to tell a bit about other intrusion possibilities via network and authentication protocols exist to stop those, but not insider attacks. Then refer the figure and explain insider attack here.}
	 In most cases, an attacker is a rogue vehicle participating in a CV application like platoon. It poses itself as a legitimate vehicle to the certificate authority (CA), like Security Credentials Management System (SCMS) \cite{whyte2013security}, thus obtaining  certificate to be a participant of a platoon. Then, it can misuse the underlying CA and launch various types of attacks that lead to collisions \cite{abdo2019application}. Fig.~\ref{figPlatoonAtk} depicts such an attack scenario where we consider a vehicle within the platoon is a rogue one and it sends false information about its position, velocity, and/or acceleration to other vehicle that it communicates with. Such type of attacks are called false data injection (FDI) attacks. A survey on different types of attacks on connected vehicles (like, message spoofing, denial of service, man in the middle, replay attack etc.) can be found in \cite{ghosal2020security}. These issues open up a serious research discussion on connected vehicle security and safety and re-calibration of the state-of-the-art control and security schemes keeping such vulnerabilities in mind from development stage.
	
	\par 
	% In the early phase of connected vehicle development cycle, the security issues were not being considered. To analyze whether the controller involved in a connected vehicle application is robust against such attacks, there is need for a \emph{customizable traffic simulation tool}. 
	There exist some state-of-the-art simulation tools for CVs \cite{lopez2018microscopic,sommer2010bidirectionally,segata2014plexe,amoozadeh2015platoon,Althoff2017a,dosovitskiy2017carla,matlabdriving}. These tools can be useful to simulate and test developed control strategies. But to the best of our knowledge, there does not exist any CV traffic simulation tool which targets reproducing  real-world CV vulnerabilities and testing control and detection strategies to counter such vulnerabilities. For example, CARLA \cite{dosovitskiy2017carla} is a widely used autonomous vehicle simulation tool which can be utilized for simulating connected vehicle scenarios. However, in Carla the vehicle dynamics model is a blackbox to the user. The dynamics of each vehicle can not be customized. CommonRoad \cite{Althoff2017a} and Matlab's Automated Driving Toolbox \cite{matlabdriving} provide the provisions  to model the vehicle dynamics as per designer's requirement. However, they lack the network protocol implementation required for CV applications. Veins \cite{sommer2010bidirectionally} integrates traffic simulator SUMO \cite{lopez2018microscopic} and network simulator OMNET++ \cite{varga2008overview} to provide a platform for simulating CV applications. Specifically, Plexe  \cite{segata2014plexe} is an extension on Veins for platoon simulation. VENTOS  \cite{amoozadeh2015platoon} was built on top of Veins to incorporate platoon management protocols. Platoon Management Protocols (PMP) govern various platoon operations and maneuvers. The leader of the platoon i.e. the front vehicle administers platoon decisions such as speed, lane changes,  merging with other platoons etc. \cite{abdo2019application}. However, VENTOS has some predefined platoon structures and dynamics of the participating vehicles. Also, the architecture of VENTOS does not support mobile attacker, for example, consider a rogue vehicle within the platoon as an attacker.
	
	\par In this paper, we present a customizable simulation tool-chain for CV traffics. The proposed tool-chain is built on top of VENTOS. 
	% As shown in Fig.~\ref{figPlatoonAtk}, we consider that a vehicle within the platoon is a rogue one and it sends false information about its position, velocity, and acceleration to other vehicle that it communicate with. Such type of attacks are called false data injection (FDI) attacks.
	It can be used to generate a sequence of false data, injection of which on sensor measurements can lead to instability in the platoon in terms of performance and safety. Such sequence of false data to be injected by the attacker vehicle is called an attack vector. The tool also provides a platform to simulate such attack on a platoon where i) each vehicle's dynamics can be customized, ii) the platoon topology can be defined. This enables us to investigate the different vulnerable situations designed for different such system dynamics and CV topologies. Customizable vehicle dynamics can be further leveraged for the development of control-theoretic attack detectors like \cite{kremer2020state, ju2020distributed,mousavinejad2019distributed}.  To this end, we state the primary contributions of this work below.
	\begin{compactenum}
		\item We propose a Satisfiability Modulo Theory (SMT)-based method that generates a sequence of false data to be added to the actual position, velocity, or acceleration of the attacker/rogue vehicle such that it either leads to collision or reduce traffic throughput by increasing the gap between the vehicles in the platoon. 
		
		\item We extend the basic VENTOS tool-flow to support external false data inputs to the application layer of the connected vehicle network in order to reproduce the vulnerability and quantify the performance degradation or understand the safety violations.
		
		\item We incorporate longitudinal vehicle dynamics in state space form in the SUMO module of VENTOS. Our proposed tool enables the user to provide parameters of the vehicle dynamics as input. 
		\item Our automated tool-chain enables a plug-and-play like interface accepting the following specifications: $(a)$ Platoon topology, $(b)$ Longitudinal vehicle dynamics and $(c)$ Attack surface variables with  corresponding false data parameters. For this specific input configuration, the tool-chain $(i)$ generates attack vectors if  exists that violates safety or performance criteria in the system model, and $(ii)$ visually simulates this  attack-effect on the provided CV and platoon models. 
		% 	Such tool will not only help in verifying robustness of large class of platoon structure against FDI attacks, but also can be used to develop control-theoretic detection systems
		
		\item We also evaluate our proposed tool for a number of platoon topologies and different attack scenarios. 
	\end{compactenum}
		\section{Background}
	\label{secBackground}
	\subsection{Platoon Topology}
	\label{subsecPlatoonTopology}
	\begin{figure}[!ht]
		% 		\captionsetup{justification=centering}
		\centering
		\begin{subfigure}{\linewidth}
			\centering			\includegraphics[width=\textwidth,keepaspectratio,clip]{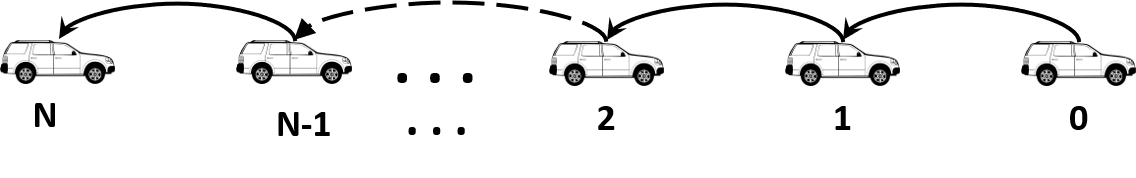}
			\vspace{-2em}
			\caption{Predecessor Following (PF)}
			\label{figPf}
		\end{subfigure}%
		\vfill
		\vspace{0.2in}
		\begin{subfigure}{\linewidth}
			\centering			\includegraphics[width=\textwidth,keepaspectratio,clip]{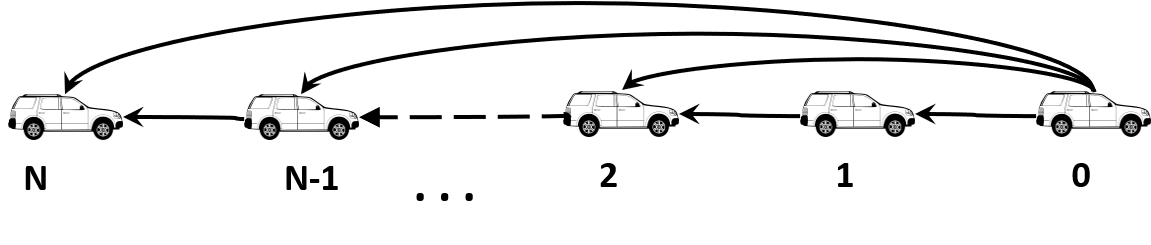}
			\vspace{-2em}
			\caption{Predecessor Leader Following (PLF)}
			\label{figPlf}
		\end{subfigure}%
		\vfill
		\vspace{0.2in}
		\begin{subfigure}{\linewidth}
			\centering			\includegraphics[width=\textwidth,keepaspectratio,clip]{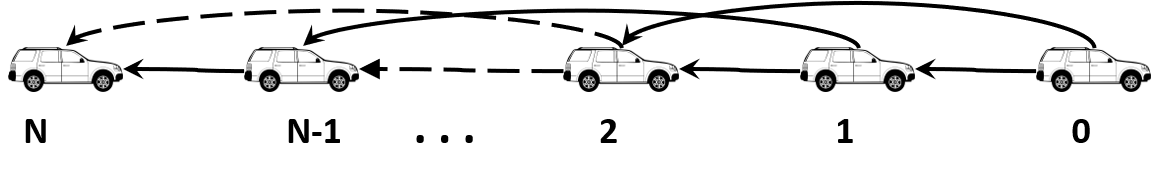}
			\vspace{-2em}
			\caption{Two Predecessor Following (TPF)}
			\label{figTpf}
		\end{subfigure}
		\vfill
		\vspace{0.2in}
		\begin{subfigure}{\linewidth}
			\centering			\includegraphics[width=\textwidth,keepaspectratio,clip]{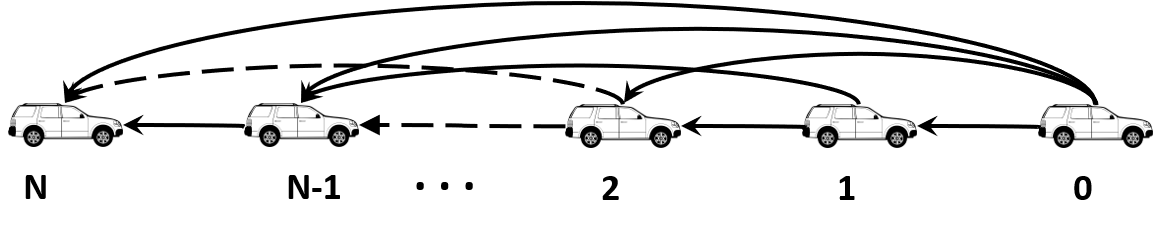}
			\vspace{-1em}
			\caption{Two
			Predecessor Leader Following (TPLF)}
			\label{figTplf}
		\end{subfigure}
		\caption{Some Platoon Topologies}
		\label{figPlatopo}
	\end{figure}
	We consider a platoon which has a total $(n+1)$ number of vehicles including a leader. As presented in Fig.~\ref{figPlatopo}, we give an index value $0$ to the platoon leader. A platoon topology defines how the vehicles participating in the platoon share their states i.e. position, speed and acceleration information among themselves either via V2X communication or radar-based measurements. In this work, we consider the following $4$ types of platoon topologies (Fig.~\ref{figPlatopo}): \begin{compactenum}[a)]
		\item Predecessor Following (PF): Each vehicle receives position, velocity and acceleration information from the vehicle just preceding it (Fig.~\ref{figPf}).
		\item Predecessor Leader Following (PLF): Each vehicle receives position, velocity and acceleration information from the vehicle just preceding it as well as from the leader (Fig.~\ref{figPlf}).
		\item Two Predecessor Following (TPF): Each vehicle receives position, velocity and acceleration information from two preceding vehicles (Fig.~\ref{figTpf}).
		\item Two Predecessor Leader Following (TPLF): Each vehicle receives position, velocity and acceleration information from two preceding vehicles as well as the leading vehicle (Fig.~\ref{figTplf}).
	\end{compactenum} 
	For input specification, we represent the platoon topology as a directed graph $G(V,E)$ \cite{zheng2015stability} where the set of nodes $V = \{\alpha_0, \alpha_1, \alpha_2, ... , \alpha_n\}$ and the set of edges $E = V\times V$. The $\alpha_i \in V$ denotes the $i$-th vehicle in the platoon where $\alpha_0$ is the platoon leader. The edge $(\alpha_i,\ \alpha_j)\in E$ denotes the information flow between $\alpha_j$ and $\alpha_i$. The information flow among the vehicles in the platoon is defined using the adjacency matrix $M \in \mathbb{R}^{(n+1)\times (n+1)}$ where
	\begin{equation}
		M[i,j]=\begin{cases}
			1, & \text{if $(\alpha_j, \alpha_i)\in E$}.\\
			0, & \text{otherwise}.
		\end{cases}
	\label{eqM}
	\end{equation}
	Here, $(\alpha_j, \alpha_i)\in E$ implies $i$-th vehicle receives data from $j$-th vehicle. The $j$-th vehicle is neighbour of $i$-th vehicle if $M[i,j] = 1$. This implies $\alpha_i$ receives information from $\alpha_j$ either via  V2X communication or radar. We define the neighbour set of $i$-th vehicle as $\mathbb{N}_i=\{j|M[i,j]=1\}$. For all the topologies in Fig.~\ref{figPlatopo}, the preceding vehicle $\alpha_j$ of each following vehicle $\alpha_{j+1}$ is a member of $\mathbb{N}_{j+1}$. Note that, in this work, though we have considered only PF, PLF, TPF, TPLF platoon topologies,  the directed graph-based formulation can support any platoon topology. 
\begin{figure*}[!ht]
\centering
\includegraphics[clip,width=0.9\linewidth]{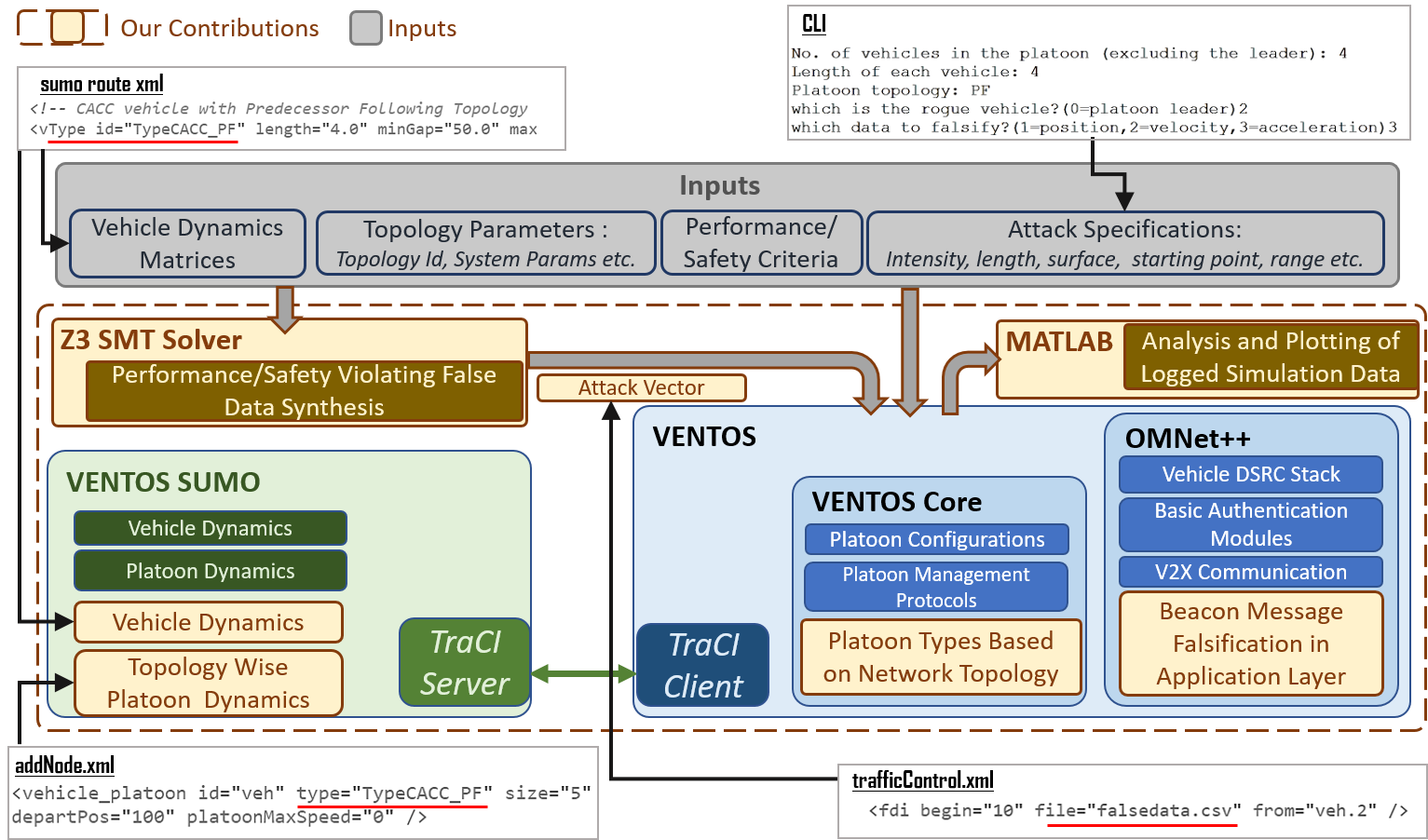}
\caption{A Framework for Evaluating Connected Vehicle Security against False Data  Injection Attacks}
\label{figToolflow}
\end{figure*}
	\subsection{Vehicle Model}
	\label{subsecVehicleModel}
\begin{figure}[H]
	    \centering
	    \includegraphics[width=\columnwidth]{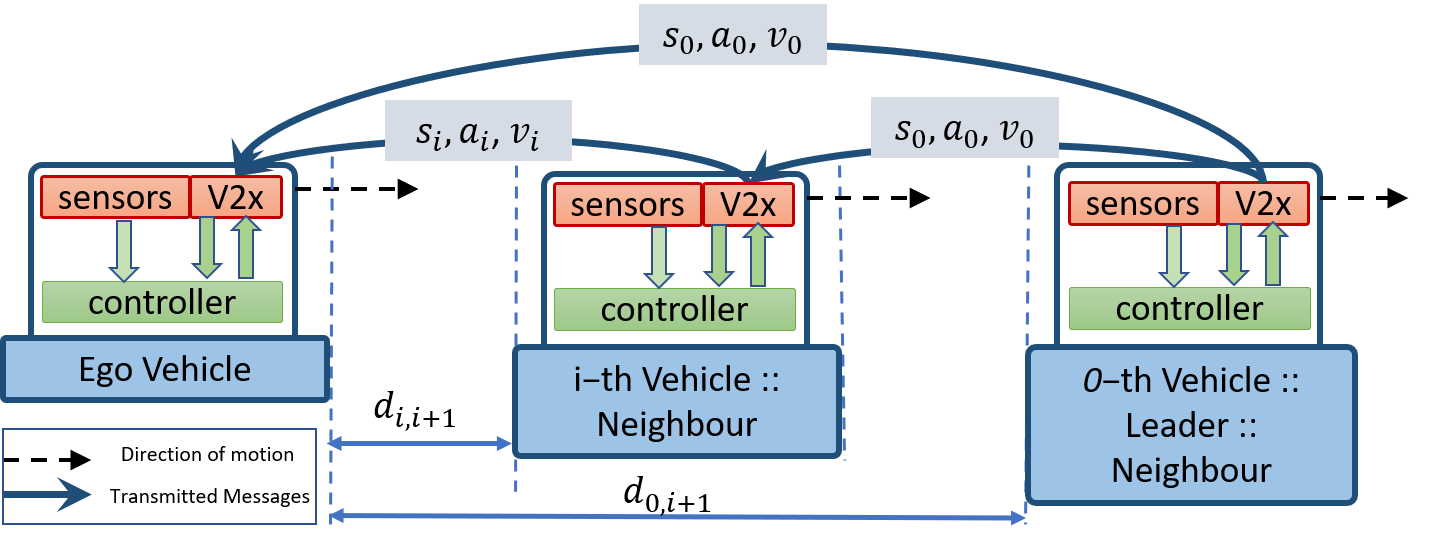}
	    \caption{Control structure of each vehicle in the platoon}
	    \label{figControlStructure}
	\end{figure}

	In this paper, we assume a homogeneous platoon i.e. dynamics of each vehicle is uniform. Fig.~\ref{figControlStructure} demonstrates a sample information flow among vehicles in a platoon. Each vehicle $\alpha_i\in V$ in the platoon is represented as a $3$-rd order continuous time state space model \cite{zheng2015stability}:
	\begin{equation}
		\dot{x}_i(t) = A_cx_i(t) + B_cu_i(t)
		\label{eqState}
	\end{equation}
	Here, the system state $x_i(t) = [s_i\ v_i\ a_i]^T$ where $s_i,\ v_i,\ a_i$ are respectively the position, velocity and acceleration of vehicle $\alpha_i$. A vehicle transmits its entire state information $x_i(t)$ to other vehicles in the platoon depending upon the platoon topology. System state transition matrices $A_c = \begin{bmatrix}
		0 & 1 & 0\\0 & 0 & 1\\0 & 0 & -\frac{1}{\tau_i}
	\end{bmatrix}$ and $B_c = [0\ 0\ \frac{1}{\tau_i}]$ where $\tau_i$ is the inertial delay of vehicle longitudinal dynamics. The linear control input $u_i(t)$ is calculated as
	\begin{align}
	\nonumber
		u_i(t) &= \sum_{j\in\mathbb{N}_i}^{}[k1(s_i - s_j - d_{i,j})+k2(v_i-v_j)+k3(a_i-a_j)]\\
		&= \sum_{j\in\mathbb{N}_i}^{}K(x_i(t) - x_j(t) - [d_{i,j}\ 0\ 0]^T)
		\label{eqControl}
	\end{align}
	Here, $d_{i,j}$ is the desired spacing between $i$-th and $j$-th vehicle including vehicle length (Fig.~\ref{figControlStructure}), and $K = [k1\ k2\ k3]$ is the gain for the linear controller. In Eq.~\ref{eqControl}, we can see the control input for each vehicle in the platoon depends on the information flow topology of the platoon. Discretizing the continuous time system in Eq.~\ref{eqState} with respect to a sampling period of $T_s$ gives the following discrete-time state space model for $i$-th vehicle in the platoon.
	\begin{align}
	\nonumber
	    x_i[k+1] &= Ax_i[k] + Bu_i[k]\\
	    u_i[k] &= \sum_{j\in\mathbb{N}_i}^{}K(x_i[k] - x_j[k] - [d_{i,j}\ 0\ 0]^T)
	    \label{eqDiscreteState}
	\end{align}
	Here, $x_i[k]$, $u_i[k]$ are respectively the state and control input of the $i$-th vehicle at $k$-th sample. A and B are the discrete-time state transition matrices.
	\subsection{Attack Model}
	\label{subsecAttackModel}
	 The inter-vehicle communication channels significantly expose the intra-vehicular network. This broadens the attack surface by allowing a remote attacker to launch an attack on the inter-vehicular network and electronic control units. Works reported in~\cite{adhikary2020skip,koley2020formal} discuss simulation of attack on intra-vehicular network that tampers sensor measurements and/or control signals. In this work, we focus on attacks that target the inter-vehicular network. We consider the attacker is a rogue vehicle in the platoon. Presenting itself as a legitimate user, it acquires certificate to be part of a platoon from certificate authority \cite{abdo2019application}. Ideally, all the vehicles in the platoon should have same velocity and the inter-vehicle space should be near optimal during stable operations. The objective of the attacker (here the rogue vehicle) would be creating instability in the platoon. This can either lead to collision i.e. the inter-vehicle space falls below the minimum criteria, or reduce traffic throughput i.e. increase the inter-vehicle space beyond the maximum limit. In this paper, we consider a false data injection (FDI) attack in which the attacker i.e. the rogue vehicle sends false information regarding its state. Consider that the $j$-th vehicle in the platoon is the attacker. It modifies its state as,
	 \begin{equation}
	     x_j^a[k] = x_j[k] + \Gamma\delta_j[k]
	     \label{eqAttack}
	 \end{equation}
	Here, $x_j^a[k]$ is the modified state of the rogue vehicle, $\delta_j[k]$ is the false data i.e. to be added to the actual state $x_j[k]$ at $k$-th sampling instance. The $3\times 1$ matrix $\Gamma$ represents the attack surface which is defined as,
	\begin{equation}
	    \Gamma[i] = \begin{cases}
			1, & \text{if $i$-th state variable is falsified}.\\
			0, & \text{otherwise}.
			\end{cases}
			\label{eqGamma}
	\end{equation}
	For example, if $\Gamma[1] = 1$ and $\Gamma[0]=\Gamma[2]=0$, then position of the rogue vehicle is modified. Such falsification will affect the control input calculation of those vehicles which receive state information from the $j$-th vehicle. Consider that the $j$-th vehicle is neighbour of $i$-th vehicle. Thus, the control input of $i$-th vehicle will be incorrectly calculated as,
	\begin{align}
	\nonumber
		u_i^a[k] = \sum_{m\in\mathbb{N}_i, m\neq j}^{}&K(x_i[k] - x_m[k] - [d_{i,m}\ 0\ 0]^T)\\
		&+ K(x_i[k] - x^a_j[k] - [d_{i,j}\ 0\ 0]^T)
		\label{eqControlAttack}
	\end{align}
	The control input $u_i^a[k]$ in the presence of the rogue vehicle can create instability in $i$-th vehicle. This would have a cascading effect on all the vehicles which have a connected network datapath from $j$-th vehicle. The attacker would inject a sequence of false data $\delta_j[k]$'s to fulfil its objective. In the next section, we present a method to synthesize a successful attack vector i.e. a sequence of false data that can hamper safety or performance of a platoon.
    \subsection{VENTOS Tool-flow}
    \label{subsecVentos}
    VENTOS is an integration of SUMO and OMNET++. Different car following models are added as CV types in its SUMO module. SUMO provides a Traffic Control Interface (TraCI), which enables TCP-based client/server interfacing of an external application to retrieve or control traffic simulation data. VENTOS connects to this TCP server in SUMO as a client. It performs network simulations for each time-step and commands SUMO to perform simulation for the same time-step. VENTOS uses a timer to call $simulationStep$ TraCI command periodically in order to advance SUMO simulation by one time-step. Each CV or RSU inserted into SUMO is mirrored in OMNET++ as a node. The connections between these nodes are set as defined by the user. The main traffic simulation part is executed in VENTOS's SUMO module and the respective network simulation for communication between CVs is done in OMNet++. 
    
    VENTOS implements a platoon management protocol (PMP) to enable different platoon maneuvers (eg. split, merge etc.). Its traffic signal control (TSC) module implements popular state-of-the-art algorithms like Fixed-time, Longest Queue First (LQF), Oldest Job First (OJF) etc. Many of the well known certification revocation list (CRL) distribution algorithms in Vehicular Public Key Infrastructure (PKI) architecture are also implemented in VENTOS. For example, RSU only i.e., only V2I communication is responsible for CRL distribution, C2C epidemic i.e., V2V and V2I communication is responsible for CRL distribution, Most Pieces Broadcast (MPB) i.e., vehicle or RSU which has maximum number of pieces is responsible for CRL distribution, etc.  
    We utilize these inbuilt features in VENTOS to implement a practical CV platooning scenario. Currently, VENTOS supports specifying new adversary nodes and modules for platoon vulnerability analysis, but it does not support an automotive attack surface. Our implementation is aimed to replicate such attack scenarios where a vehicle that joined the platoon is rogue and injects false data to cause unsafe situations. VENTOS supports specifying new vehicle types but its control algorithm does not consider different platoon network topologies. We have parameterized this longitudinal vehicle dynamics integration along with different type definitions for different platoon topology types as discussed below.
    
\section{Proposed Tool Framework}
\label{secProposedMethod}
Our proposed tool chain has primarily two parts: attack vector generation and attack simulation. We now discuss the input specification of the tool chain followed by a detailed discussion of these two parts.

\subsection{Tool Input Specifications: }
\label{subsecInputSpec}
To enable a customizable system configuration, we consider the following inputs as shown in the grey box in Fig.~\ref{figToolflow}. 
\begin{itemize}
\item \textit{Vehicle Dynamics Matrices and Parameters :} The user can provide the discrete-time state-space matrices $A$, $B$, the gain $K$ along with the sampling period $T_s$ of the controller to define longitudinal vehicle dynamics in ACC mode. Currently, the tool-chain supports homogeneous modeling of CV platoons following Eq.~\ref{eqState}. The tool also supports command line input for other vehicle parameters like length, desired spacing $d$ etc.
\item \textit{Topology Parameters :} The user can provide as input the number of vehicles $n$ (excluding leader) in the platoon and the topology type. The adjacency matrix $M$ will be generated for the CV platoon topology as mentioned in Eq.~\ref{eqM}. Currently, our implementation supports $4$ platoon topology types i.e., PF, PLF, TPF, TPLF (see Sec.~\ref{subsecPlatoonTopology}). Using these information the platoon dynamics and control equations are formed following Eq.~\ref{eqDiscreteState} and Eq.~\ref{eqControl}. 

\item \textit{Performance/Safety Criteria:}  Typically, interesting attack vectors for FDI  security evaluation would be those which lead to safety violation or performance violation. Both attack $types$ are supported by the tool. The target performance or safety property is considered as an input to the tool. Safety properties in this case are defined by the minimum distance $d_{min}$ between any $2$ consecutive vehicles to avoid collision and performance criteria is defined by maximum distance $d_{max}$ between any $2$ consecutive vehicles in the platoon exceeding which may reduce traffic throughput. The system dynamics of the CV platoon should satisfy these properties during normal operations (i.e, when the false data is not injected). A false data sequence is generated with a goal to violate these safety or performance criteria. 
% Currently, the tool-chain takes command line input to specify, $(i)$ the type of property ($safety$ or $performance$) to be violated by the attacker, and $(ii)$ the safety or performance boundary to be violated.
\item \textit{Attack Specifications :}
The attack specifications include $(i)$ the attack surface matrix $\Gamma$ that denotes which among position, velocity and acceleration is under attack, $(ii)$ which one of the platoon vehicles is the attacker vehicle $p$ (considering leader is indexed with $0$), $(iii)$ sampling instance $s$ when the attack will start, $(iv)$ duration $T$ of attack, and $(v)$ the range $[-\theta, \theta]$ that each attack value i.e. the false data falls into. 
\item \textit{Leader's velocity profile :} Ideally, all the vehicles in the platoon should follow the leader and maintain the same velocity as leader. Therefore, it is also needed to give the time stamped velocity profile $v$ (a vector of speed  values over a time window) of the leader as input to the tool. Leader's dynamics will change based on $v$. With these command line inputs the attack vector is first synthesized and then fed in to the simulation engine.
\end{itemize}

\subsection{Attack Vector Generation}
	\label{subsecAttackGeneration}
	In this section, we discuss a Satisfiability Modulo Theory (SMT)-based method to synthesize successful attack vectors. Particularly, we have used Z3 \cite{de2008z3}, an SMT solver for this purpose. Attack vector synthesis for platoon using SMT was first introduced in \cite{kremer2020state}. However, the authors in \cite{kremer2020state} used only PF platoon topology with only $3$ vehicles. Whereas, our proposed method presented in Algorithm~\ref{algAttackSyn} can be used to generate attack vectors for any platoon topology with any number of vehicles. 
% 	\par Algorithm~\ref{algAttackSyn} takes as input i) the discrete-time state transition matrices $A$ and $B$, ii) controller gain $K$, iii) sampling period $T_s$ of the controller, iv) number of vehicles $n$ in the platoon other than platoon leader, v) adjacency matrix $M$ (Eq.~\ref{eqM}) that represents the platoon topology, vi) attack surface matrix $\Gamma$ (Eq.~\ref{eqGamma}) that denotes which among position, velocity and acceleration is under attack, vii) attacker i.e. rogue vehicle's index $p$ in the platoon considering leader is indexed with $0$, viii) attack duration $T$ in terms of samples, ix) attack type $type$ that indicates whether we are synthesizing safety violating attack vector or performance violating one, x) sampling instance $s$ when attack begins, xi) the range $[-\theta, \theta]$ that each attack value i.e. the false data falls into, xii) desired spacing $d$ between any $2$ consecutive vehicles in the platoon, xiii) safety limit $d_{min}$ i.e. the minimum space between any $2$ consecutive vehicles in the platoon to avoid collision, xiv) performance limit $d_{max}$ i.e. the maximum space between any $2$ consecutive vehicles in the platoon exceeding which leads to reduced traffic throughput, xv) velocity profile $v$ of the platoon leader which is to be followed by all the other $n$ vehicles in the platoon, and xvi) initial velocity $v_{init}$ of each vehicle in the platoon.
	\begin{algorithm}[!ht]
		\caption{Attack Vector Synthesis for a platoon}
		\label{algAttackSyn}
		\begin{algorithmic}[1]
			\Require{Discrete-time state transition matrices $A$ and $B$, controller gain $K$, sampling period $T_s$, number of vehicles $n$ in the platoon other than leader, adjacency matrix $M$ attack surface matrix $\Gamma$, attacker vehicle number $p$, attack duration $T$, attack type $type$, attack onset $s$, attack range [$-\theta$, $\theta$], desired spacing $d$ between consecutive vehicles, safety limit $d_{min}$, performance limit $d_{max}$}, velocity profile of the leader $v_0$, initial velocity of each vehicle $v_{init}$
			\Ensure{Attack vector $\mathcal{A}$(if it exists, otherwise NULL)}
			\Function{AttVecSyn}{$A$, $B$, $K$, $T_s$, $n$, $M$, $\Gamma$, $p$, $T$, $type$, $s$, $|\theta|$, $d$, $d_{}min$, $d_{max}$, $v_0$, $v_{init}$}
			\For{$i=0$ to $n$} \Comment{Initialization}\label{algInitStart}
			\State $s_i\gets(N-i+1)*d;\quad v_i\gets v_{init};\quad a_i\gets 0;$
			\State $x_i[0] \gets [s_i\ v_i\ a_i]^T$
			\EndFor\label{algInitEnd}
			\For{$k=1$ to $s+T$}\label{algTimeStarts}
			\State $v_{prev}\gets v_0$;\label{algLeaderPrevVel}
			\State $v_0\gets v(k);\ a_0\gets\frac{v_0 - v_{prev}}{T_s};$ \label{algLeaderVelAcc}
			\State $s_0\gets s_0 + (v_{prev}T_s) + (0.5a_0T_s^2)$\label{algLeaderPos}
			\State $x_0[k] \gets [s_0\ v_0\ a_0]^T$ \label{algLeaderState}
			\For{$i=1$ to $n$}\label{algVehLoopStart}
			\State $u_i[k-1]\gets 0$ \label{algControlInit}
			\For{$j=1$ to $i-1$}\label{algPrecedStarts}
			\If{$M[i,j] == 1\ \&\ j==p\ \&\ k\geq s$}\label{algCheckIfToAttack}
			\State $\delta[k-s]\gets nondet();$\label{algNotDet}
			\State $d_{i,j}\gets (i-j)\times d$\label{algComputeAttackD}
			\State $u_i[k-1]\gets u_i[k-1] - K(x_i[k-1] - (x_j[k-1]+\Gamma\delta[k-s]) - [d_{i,j}\ 0\ 0]^T)$\label{algComputeAttackControl}
			\EndIf\label{algIfAttackCheckEnds}
			\If{$M[i,j] == 1$}\label{algIfNeighbourCheck}
			\State $d_{i,j}\gets (i-j)\times d$\label{algComputeD}
			\State $u_i[k-1]\gets u_i[k-1] - K(x_i[k-1] - x_j[k-1] - [d_{i,j}\ 0\ 0]^T)$\label{algComputeControl}
			\EndIf\label{algNeighbourCheckEnds}
			\EndFor \label{algPrecedEnds}
			\State $x_i[k]\gets Ax_i[k-1] + Bu_i[k-1]$ \label{algVehStateUpdate}
			\State $e_i[k]\gets s_{i-1} - s_i$ \label{algError}
			\EndFor \label{algVehLoopEnds}
			\EndFor\label{algTimeEnds}
			\If{$type == safety$} \Comment{safety violation}\label{algIfSafety}
% 			\State $\mathbb{A}_{safety}\gets\textbf{assert}( \forall k\in[s,T], \forall i\in[p+,n], (\bigwedge|\delta[k]|<\theta)\to(\bigvee e_i[k]<d_{min}))$ \label{algAssertionSafety}
            \State $\mathbb{A}_{safety}\gets\textbf{assert}( \forall k\in[s,T], \forall i\in[1,p]\wedge [p+2,n], (\bigwedge|\delta[k]|<\theta)\to(\bigvee e_i[k]<d_{min}))\wedge(e_{p+1}[k]>=d_{min}\wedge e_{p+1}[k]<=d_{max}) $ \label{algAssertionSafety}
			\Else \Comment{performance violation}\label{algIfPerf}
			\State $\mathbb{A}_{perf}\gets\textbf{assert}( \forall k\in[s,T], \forall i\in[p,n], (\bigwedge|\delta[k]|<\theta)\to(\bigvee e_i[k]>d_{max}))$\label{algAssertionPerf}
			\EndIf			\label{algIfTypeEnds}
			\If {$\mathbb{A}_{safety}$ \textbf{or} $\mathbb{A}_{perf}$ is {\em valid}}\label{algIfAssertStarts}
			\State \Return $\mathcal{A} \gets
			\begin{bmatrix}
				\delta[1] & \cdots &  \delta[T]\\
			\end{bmatrix}$;\label{algAssertionCheck}
			\Else\label{algElseAssertStarts}
			\State\Return NULL;\label{algReturnNull}
			\EndIf			\label{algIfAssertEnds}
			\EndFunction
		\end{algorithmic}
	\end{algorithm}
		\par Algorithm~\ref{algAttackSyn} takes as input all those parameters that are specified in Sec.~\ref{subsecInputSpec}. We initialize position, velocity, acceleration of each vehicle including leader in lines~\ref{algInitStart}-\ref{algInitEnd} of Algorithm~\ref{algAttackSyn}. We consider that initially the platoon is in a stable state i.e. each vehicle has same velocity $v_{init}$ and the distance between any $2$ consecutive vehicles in the platoon is optimal i.e. $d$. From line~\ref{algTimeStarts} to line~\ref{algTimeEnds}, we update the states of each of the $(n+1)$ vehicles for $(s+T)$ sampling instances. The velocity profile of the leader has been given as input using which we update the states of the platoon leader in lines~\ref{algLeaderPrevVel}-\ref{algLeaderState}. The states of the $n$ following vehicles are updated in lines~\ref{algVehLoopStart}-\ref{algVehLoopEnds}. The control input for each $i$-th vehicle is computed in lines~\ref{algControlInit} to \ref{algIfAttackCheckEnds}. In line~\ref{algCheckIfToAttack}, we are checking whether the $j$-th neighbour of $i$-th vehicle is the rogue one and if the current iteration falls within the attack window $[s, s+T]$. Accordingly, we create a vector $\delta$ that signifies the false data sequence to be added by the attacker and populate it \emph{non-deterministically} in each iteration during the attack window (line~\ref{algNotDet}). These non-deterministic values are then added to the communicated states of $j$-th vehicle using which the control input of $i$-th vehicle is updated following Eq.~\ref{eqControlAttack} in line~\ref{algComputeAttackControl}. If the $j$-th neighbour of vehicle $\alpha_i$ is not the rogue one or the current iteration does not fall into the attack window $[s, s+T]$, the control input of $\alpha_i$ is computed using Eq.~\ref{eqControl} (line~\ref{algComputeControl}). The states of each vehicle $\alpha_i\in V$ are updated in line~\ref{algVehStateUpdate}. We compute an error term $e_i$ for every vehicle $\alpha_i\in V$ that signifies the distance of $i$-th vehicle from its preceding vehicle i.e. $(i-1)$-th vehicle.
	\par By \emph{a successful safety violating} attack vector we mean the false data to be added by the rogue vehicle must be within the given range $[-\theta, \theta]$ and as an effect, any two consecutive vehicles other than the vehicles in vicinity of the rogue one (so that the rogue vehicle itself does not face any accident or has enough time to leave the platoon after broadcasting falsified data) become very close i.e. below specified safe gap $d_{min}$ that can lead to collision. This is captured in the assertion $\mathbb{A}_{safety}$ (line~\ref{algAssertionSafety}). By \emph{a successful performance violating} attack vector we mean the false data to be added by the rogue vehicle must be within the given range $[-\theta, \theta]$ and also the sequence of such false data can increase the distance between any two consecutive vehicles int he platoon that reduces the traffic throughput. Depending upon whether user wants safety violating attack vector (i.e. $type = safety$) or performance violating attack vector (i.e. $type = perf$), either $\mathbb{A}_{safety}$ or $\mathbb{A}_{perf}$ will be will be given as input to the SMT solver (line~\ref{algIfAssertStarts}). If either of the assertions is satisfied, the algorithm gives as output a successful attack vector of length $T$ (line~\ref{algAssertionCheck}). Else, it returns NULL (line~\ref{algReturnNull}) which signifies that the safety or performance of the system can not be violated by any attack of duration $T$ samples.
\subsection{Attack Vector Simulation}
\label{subsecSimulationTool}
In previous section, we have proposed an SMT-based FDI attack synthesis algorithm for different longitudinal vehicle dynamics and platoon configurations (The top left box inside the dotted rectangle in Fig.~\ref{figToolflow}). In this section, we discuss about the next part of the proposed tool-chain that is built on top of VENTOS in order to analyse the effect of the attack on the CV platoons. In Sec.~\ref{subsecVentos} we have briefly discussed how VENTOS collaboratively runs OMNet++ network simulator and SUMO traffic simulator equipped with the state-of-the-art features. We now discuss the further developments done on top of its existing tool-flow for analysing security vulnerabilities of CV platoons. 

\textbf{Simulation Module Inputs: }
To enable a customizable system configuration, we consider the following inputs as shown in the grey box at the top in Fig.~\ref{figToolflow}. $(i)$\textit{Vehicle Dynamics Matrices :} The user can provide the state-space matrices to define longitudinal vehicle dynamics in ACC or CACC mode. Currently, the tool-chain supports homogeneous modeling of CV platoons.
$(ii)$ \textit{Topology Parameters :} As mentioned earlier, the topology type along with the number of vehicles are taken as inputs. Accordingly the adjacency matrix $M$ is built and control input is calculated (Eq.~\ref{eqControl}). Currently it supports $4$ platoon topology type inputs i.e., PF, PLF, TPF, TPLF (See Sec.~\ref{subsecPlatoonTopology}) 
% \st{via xml (existing addNode.xml in VENTOS SUMO)}. 
% . The user can input other topology parameters like longitudinal inertial delay for a CV due to the platoon topology $\tau$ \sd{what is this notation suddenly}. 
$(iii)$ \textit{Attack Vector and Other Specifications:} The attack vector synthesized with Algorithm~\ref{algAttackSyn} is input to this extended VENTOS simulation interface using a comma-separated value (CSV) file which is generated by the attack vector synthesis module.
% format file auto generated by our tool.
% \st{as csv (via trafficControl.xml)}
 Along with that, other attack specifications like the rogue vehicle (that broadcasts the falsified data) index in the platoon, the starting instance of the attack, etc. and the leader vehicle's velocity profile are also input to the simulation platform in order to  visualize the attack-effect. 
%  Along with these, we also provide the leader vehicle velocity profile as in a similar format.

\textbf{Design Details :}
All these inputs as discussed above are provided to the VENTOS-based extended simulation engine. At first, the system specifications and platoon configurations are parsed via command line interface (CLI) and respective CSV or XML files (refer Fig.~\ref{figToolflow}). These configurations contain vehicle dynamics, platoon parameters etc. as explained earlier. These are then passed via TraCI to the VENTOS's SUMO module. The adjacency matrix $M$ is formed according to the input topology (Eq.~\ref{eqM}). Using this, SUMO formulates the dynamical equations as shown in Eq.~\ref{eqDiscreteState}. The control input is also calculated following Eq.~\ref{eqControl}. This closed-loop dynamics is then visually simulated for a certain input time horizon.

The attack vector is input along with the attack surface details to VENTOS (see \texttt{trafficControl.xml} in Fig.~\ref{figToolflow}). The attack surface specifies the $\Gamma$ matrix as mentioned in Eq.~\ref{eqGamma} along with the rogue vehicle index ($j$ as shown in Sec.~\ref{subsecAttackModel}). While sharing its system states with its neighbouring vehicle (for a certain topology), the rogue vehicle is supposed to add these synthesized false data with the original state information, as shown in Eq.~\ref{eqAttack}. To achieve this, the parsed false data are passed to the application layer. While a wave short message (wsm) instance is generated for broadcasting, the state information packed in it is falsified using the attack vector input. The neighbouring nodes receive a falsified data which are sent to the corresponding vehicle nodes in SUMO via TraCI. This process is repeated in each time-step as long as the attack vector is injected. These victim vehicle nodes in SUMO calculates their control action following Eq.~\ref{eqControlAttack}. This causes the violations of expected performance or safety properties that are targeted by the attackers which can be visualized using simulation. We log the CV platoon states under attack utilizing logging facility provided in VENTOS and analyse them in Matlab simulation in order to understand how attack points and surfaces can be chosen such that the CV platoon is most vulnerable. Using such quantified attack scenarios, we can propose better control and detection schemes in order to make the CV platoons attack resilient. In the next section, we discuss such attack vector generation and visualize their effects.

\section{Experimental Results}
\label{secExperiment}
\begin{figure}[!ht]
		\centering
		\begin{subfigure}{0.5\columnwidth}
			\centering			\includegraphics[width=\linewidth,keepaspectratio,clip]{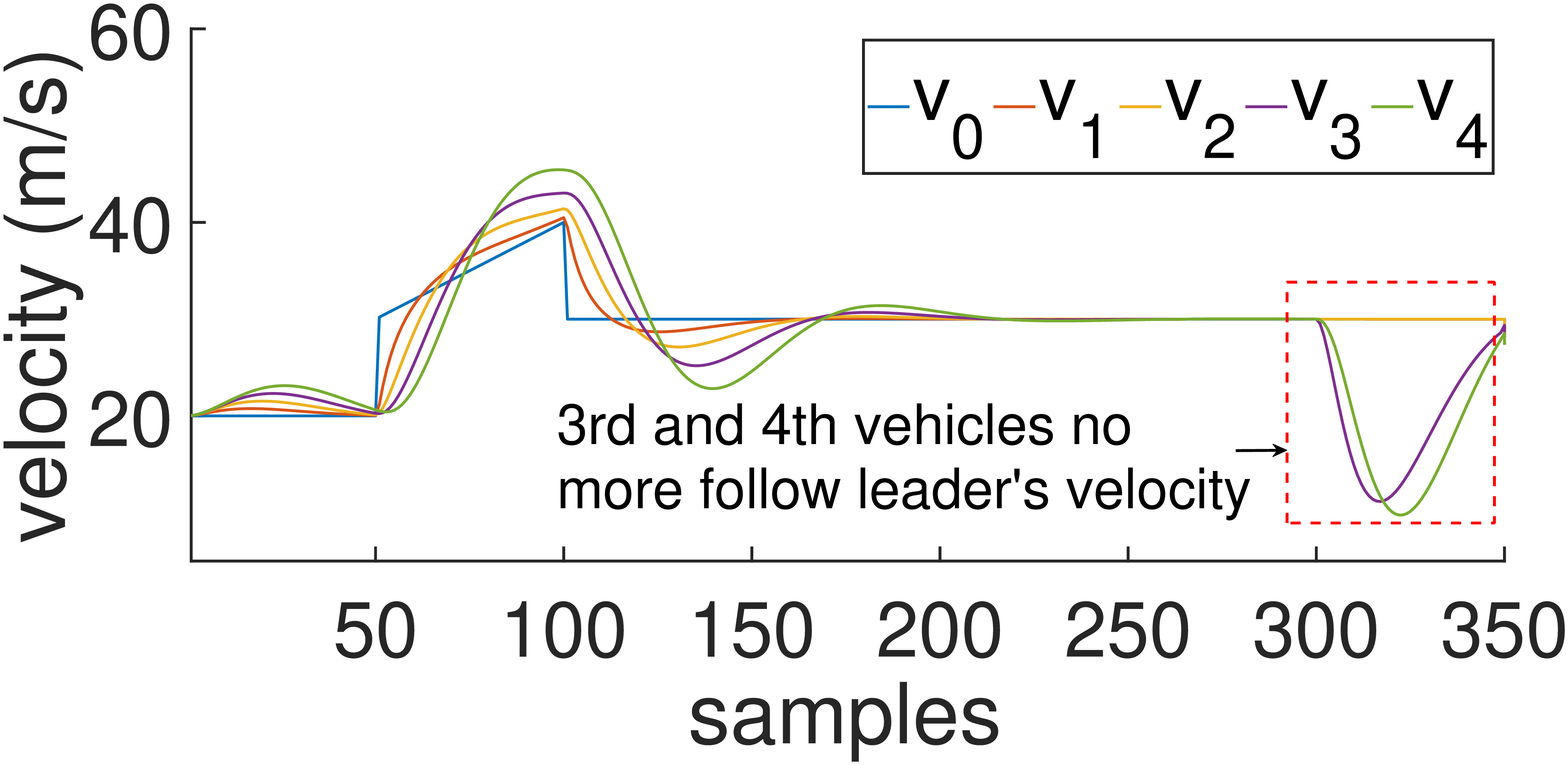}
			\caption{PF: velocity under attack}
			\label{figPfVel}
		\end{subfigure}%
		\hfill
		\begin{subfigure}{0.5\columnwidth}
			\centering			\includegraphics[width=\linewidth,keepaspectratio,clip]{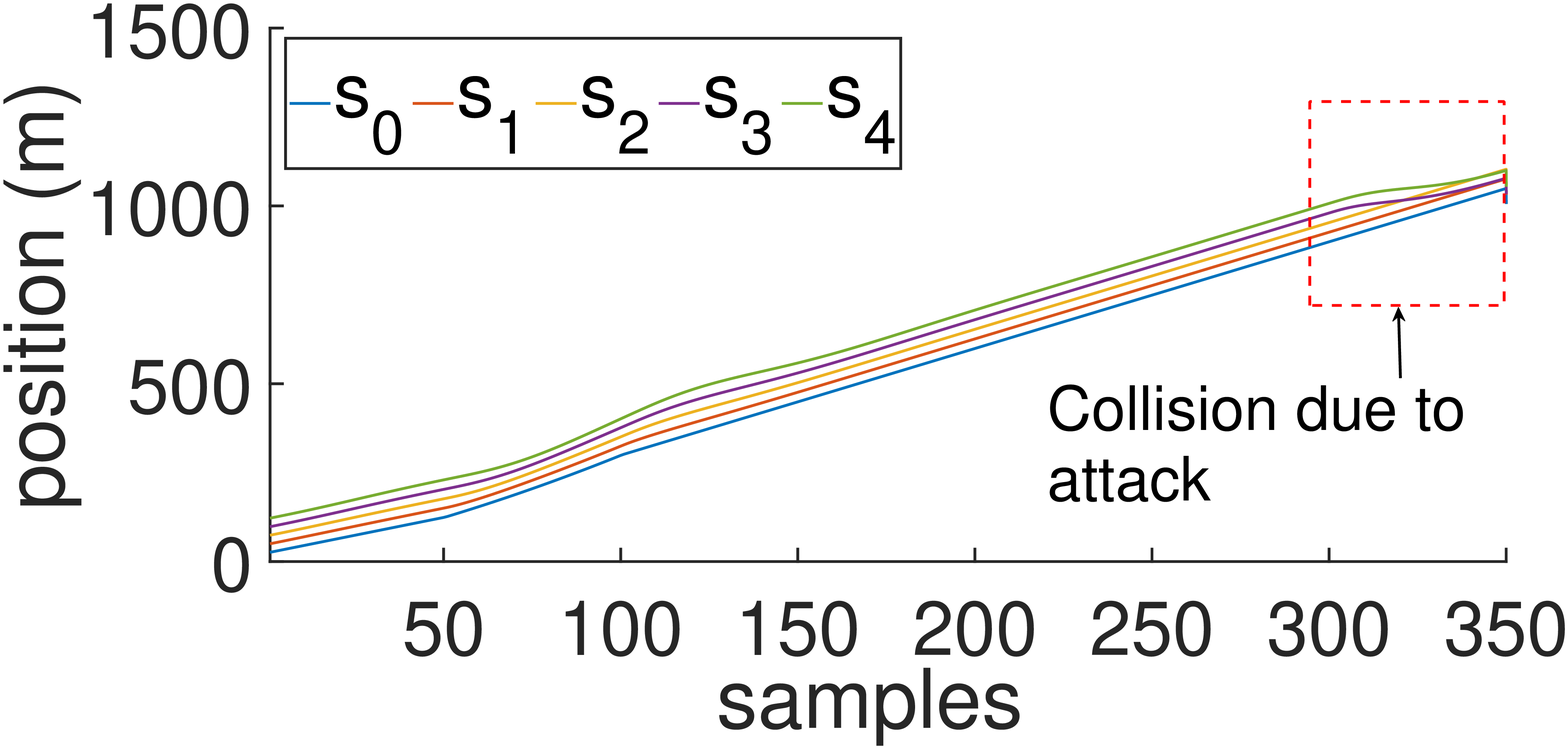}
			\caption{PF: position under attack}
			\label{figPfPos}
		\end{subfigure}%
		\vspace{0.2in}
		\begin{subfigure}{0.5\columnwidth}
			\centering			\includegraphics[width=\linewidth,keepaspectratio,clip]{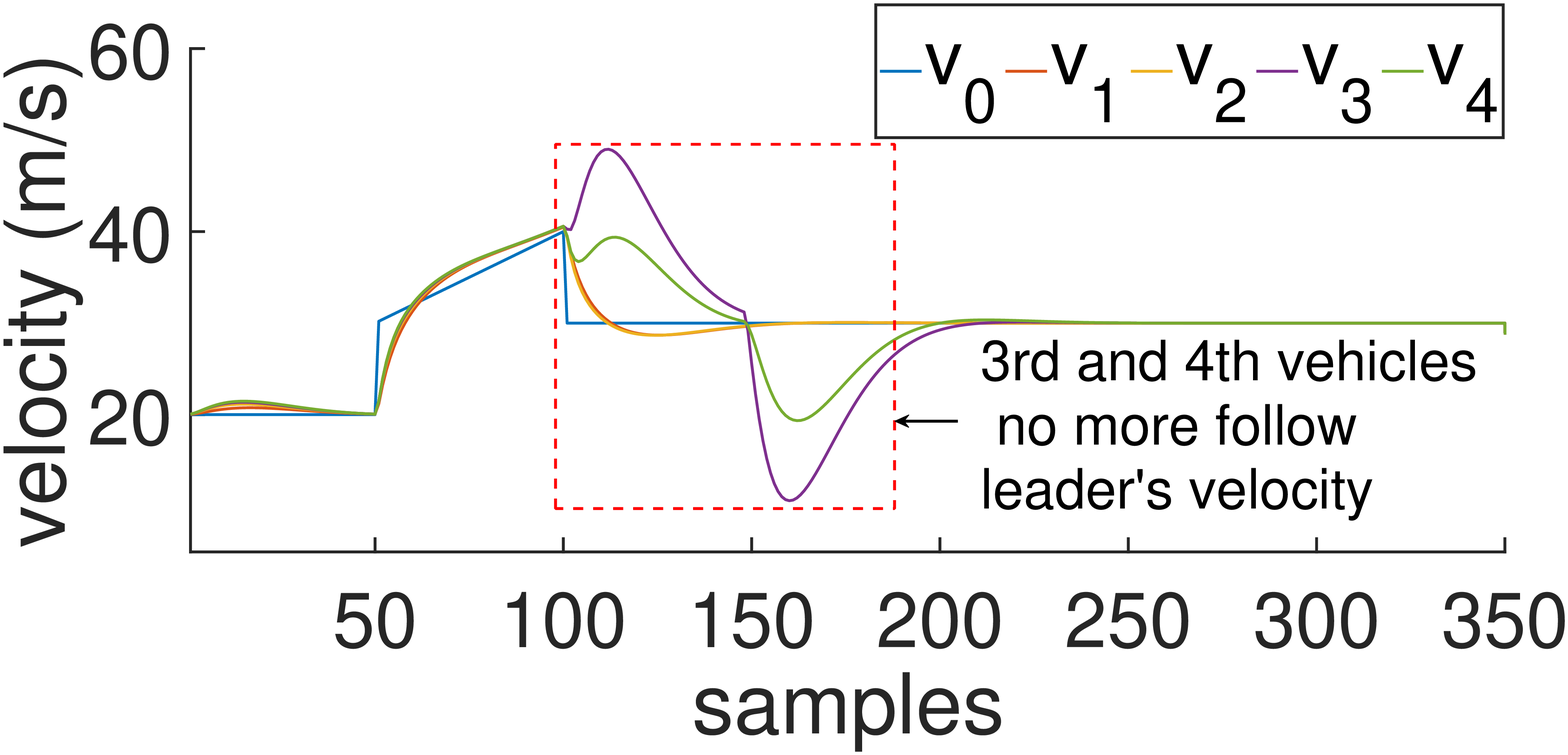}
			\caption{PLF: velocity under attack}
			\label{figPlfVel}
		\end{subfigure}%
		\hfill
		\begin{subfigure}{0.5\columnwidth}
			\centering			\includegraphics[width=\linewidth,keepaspectratio,clip]{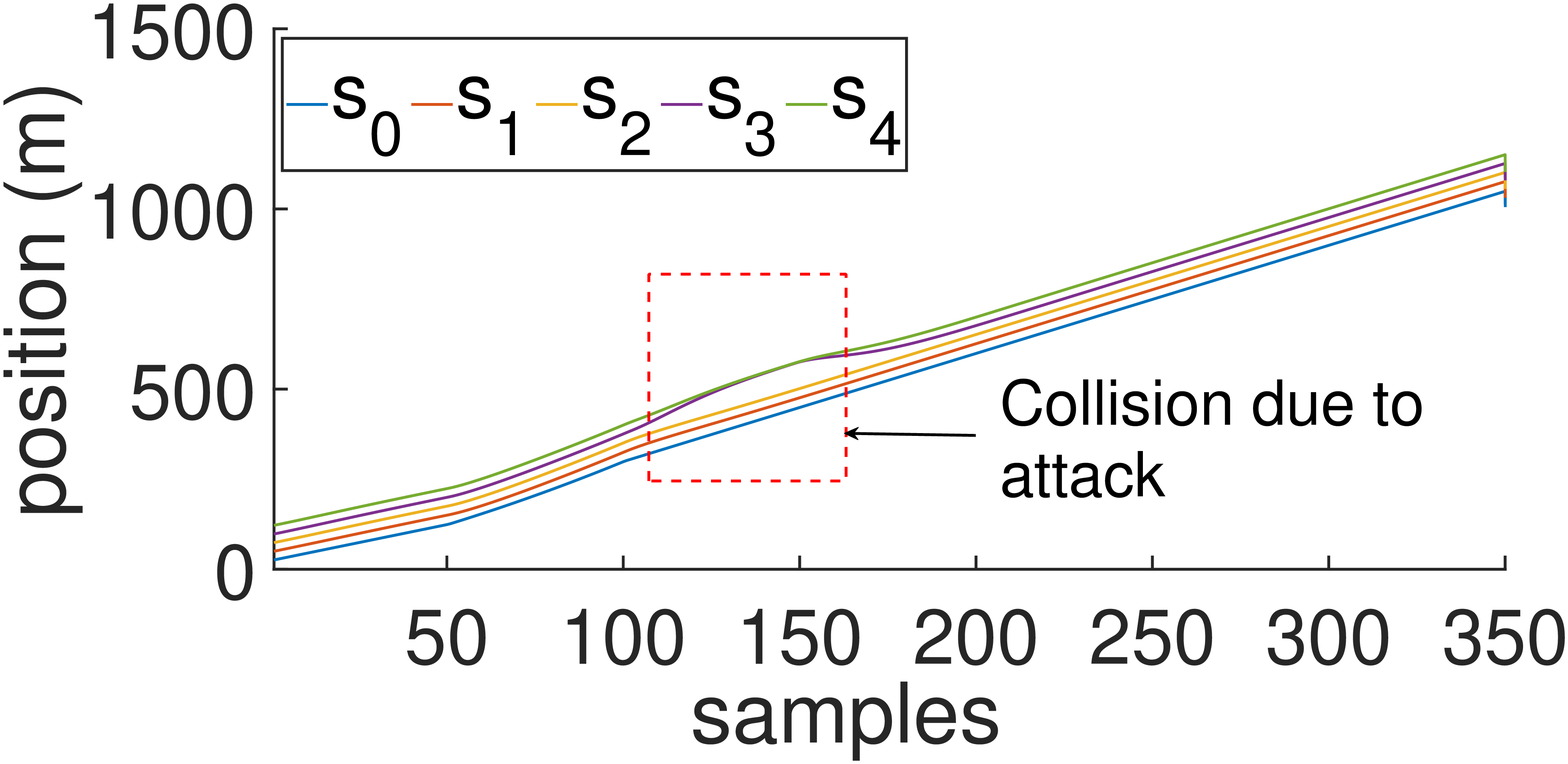}
            \caption{PLF: position under attack}
			\label{figPlfPos}
		\end{subfigure}%
		\vspace{0.2in}
		\begin{subfigure}{0.5\columnwidth}
			\centering			\includegraphics[width=\linewidth,keepaspectratio,clip]{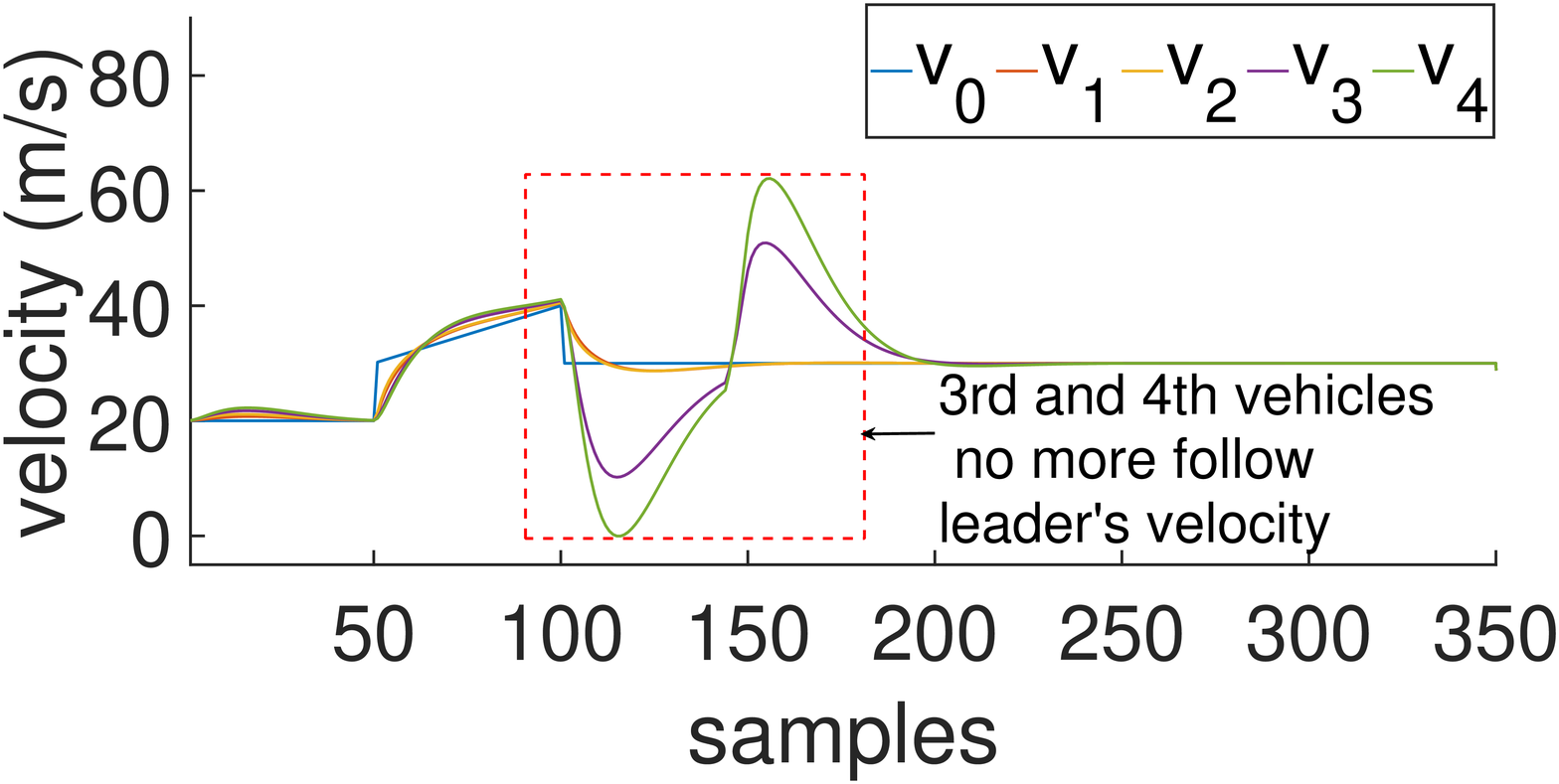}
			\caption{TPF: velocity under attack}
			\label{figTpfVel}
		\end{subfigure}%
		\hfill
		\begin{subfigure}{0.5\columnwidth}
			\centering			\includegraphics[width=\linewidth,keepaspectratio,clip]{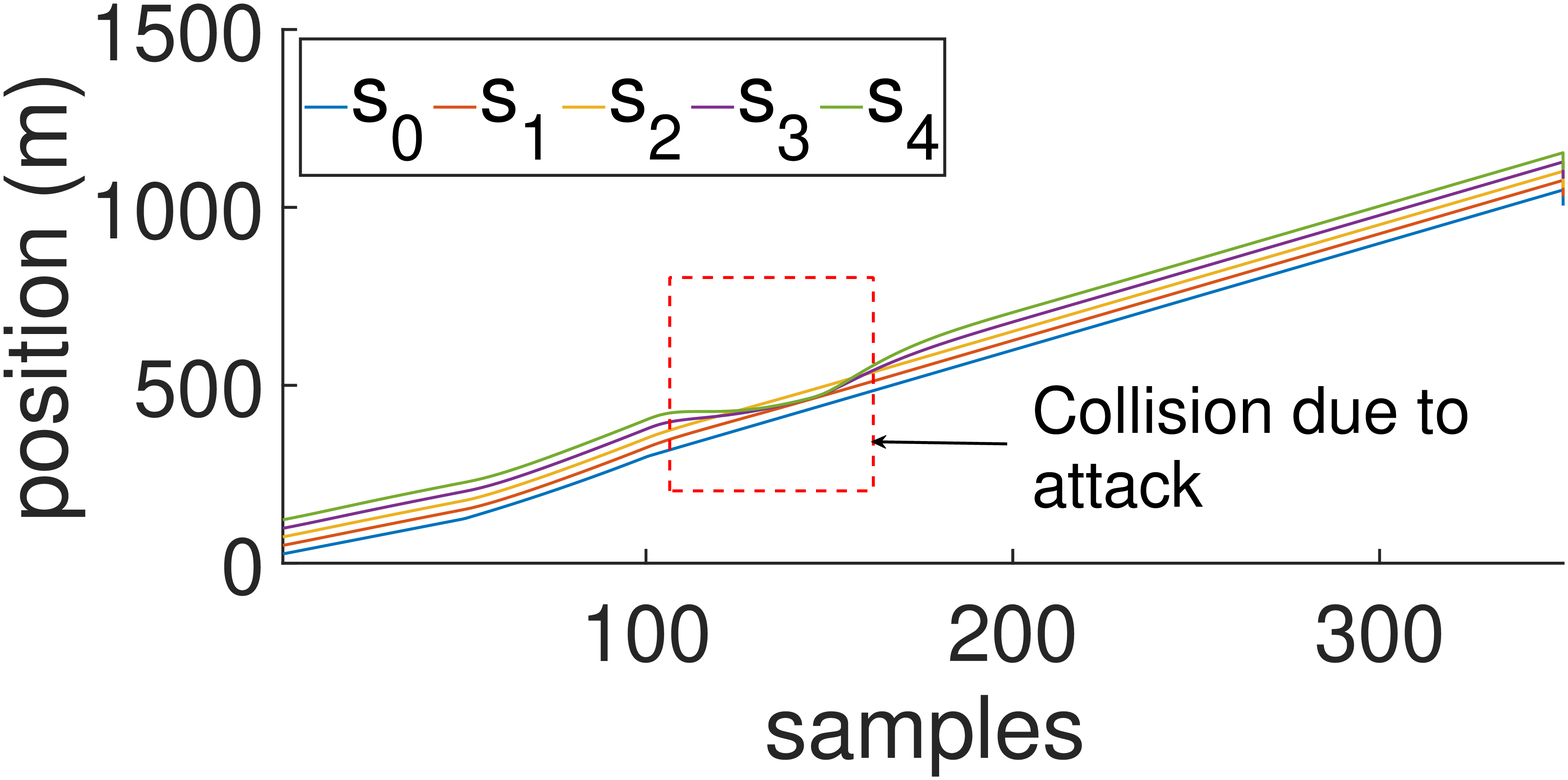}
			\caption{TPF: position under attack}
			\label{figTpfPos}
		\end{subfigure}%
		\vspace{0.2in}
		\begin{subfigure}{0.5\columnwidth}
			\centering			\includegraphics[width=\linewidth,keepaspectratio,clip]{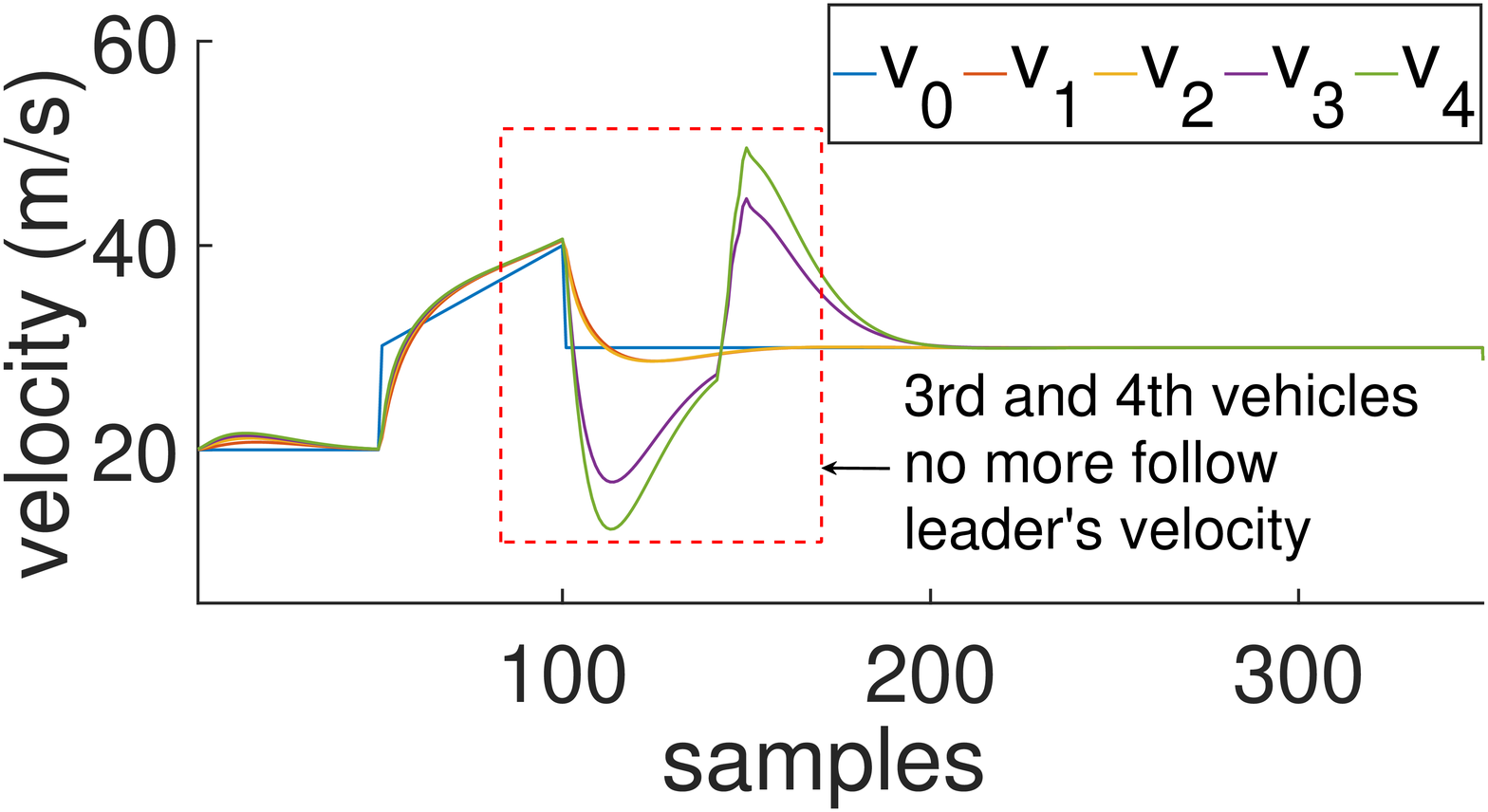}
			\caption{TPLF: velocity under attack}
			\label{figTplfVel}
		\end{subfigure}%
		\hfill
		\begin{subfigure}{0.5\columnwidth}
			\centering			\includegraphics[width=\linewidth,keepaspectratio,clip]{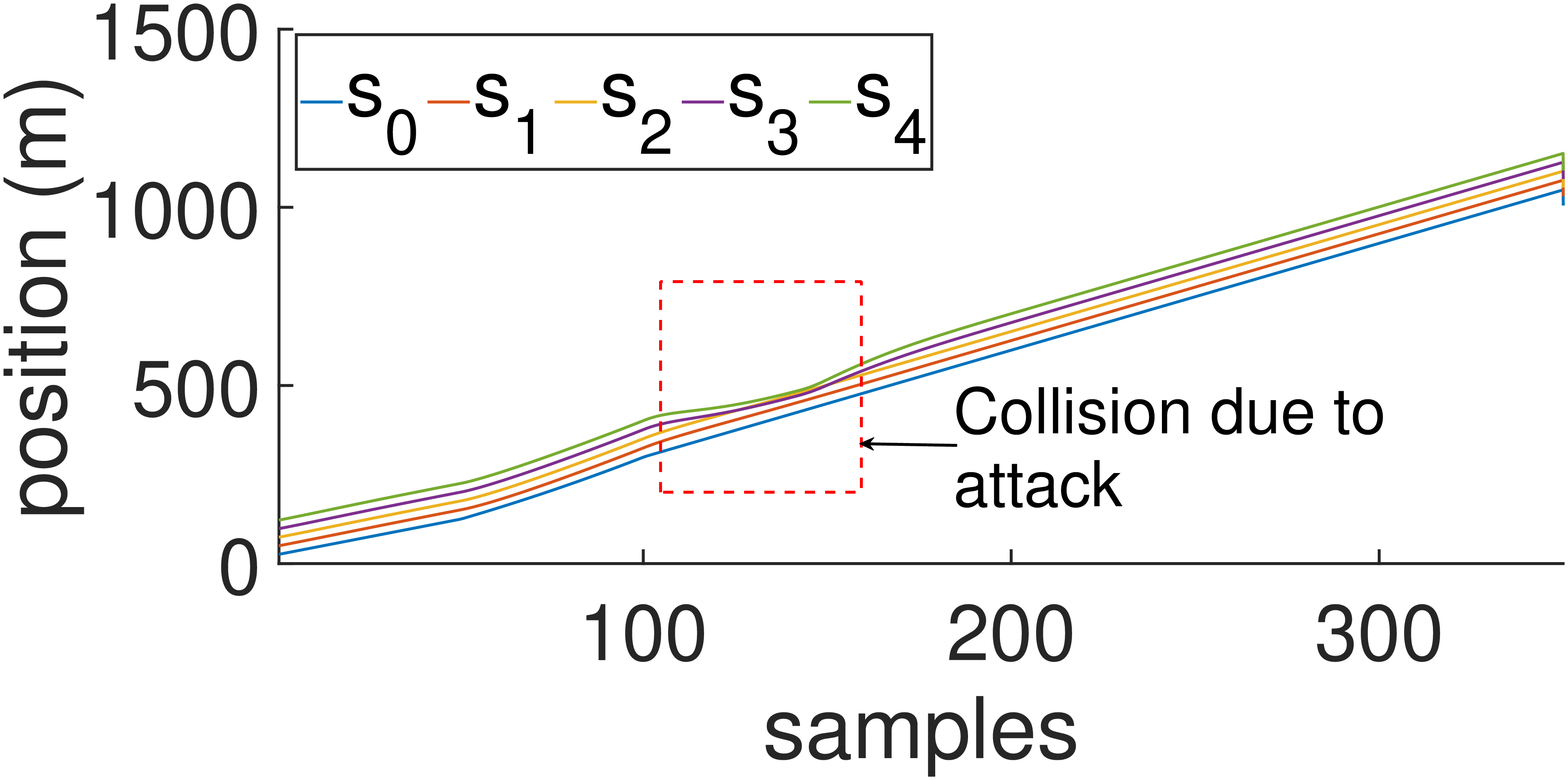}
			\caption{TPLF: velocity under attack}
			\label{figTplfPos}
		\end{subfigure}%
		\caption{Effect of FDI Attack on Platoon Topologies}
		\label{figAttackOnPlatoon}
	\end{figure}

Our automated tool-chain evaluates the security of CVs against FDI attacks. For our experiments, we have considered the continuous-time system matrices $A_c$ and $B_c$ with the parameter $\tau=0.5$, controller gain $K=[1\ 2\ 1]^T$, and desired spacing (including vehicle's length) $d = 20 m$ as given in \cite{zheng2015stability}. We discretize the linear system (Eq.~\ref{eqState}) with a sampling period $T_s = 0.1 s$. We simulate the platoon with $n=4$ i.e., there  are 4 vehicles following the platoon leader. We consider $4$ types of topology (PF, PLF, TPF, TPLF) as discussed earlier in Sec.~\ref{subsecSimulationTool}. From the input topology type (provided via \texttt{addNode.xml} as shown in Fig.~\ref{figToolflow}), the corresponding control function is chosen (as defined in \texttt{route.xml} files in SUMO, see Fig.~\ref{figToolflow}). We consider that the $2$-nd vehicle in the platoon i.e., $\alpha_2$ is the rogue one and it modifies its acceleration information while broadcasting its current states to other vehicles i.e. $\Gamma = [0\ 0\ 1]^T$. We specify the safety and performance limits as $d_{min} = 50$ and $d_{max} = 60$ (via CLI as shown in Fig.~\ref{figToolflow}). The velocity profile $v$ of the leader is input via CSV and is taken from~\cite{zheng2015stability}. The synthesized $safety$ or $performance$ violating false data sequence is dumped into a CSV file and fed as an input to our simulation module. The effect of safety violating attack vectors on different types of platoon topology is presented in Fig.~\ref{figAttackOnPlatoon}.
\par Initially, we set the range of the false data as $[-\theta, \theta] = [-50, 50]$. The effect of a safety violating attack vector of length $T=50$, generated with this range on PF platoon topology, can be seen in Fig.~\ref{figPfVel} and \ref{figPfPos}. We can observe that velocity of vehicles $\alpha_3$ and $\alpha_4$ stop following the leader's velocity due to the attack (Fig.~\ref{figPfVel}). This leads to a collision i.e., the positions of  $\alpha_3$ and $\alpha_4$ overlap in Fig.~\ref{figPfPos}. However, the attack vector generation module could not return any safety violating attack vector of length $T=50$ and range $[-\theta, \theta] = [-50, 50]$ for PLF, TPF, and TPLF topologies. 
% This implies that this attack specification, the platoon can not be made unsafe . 
Thereafter, by increasing the attack range from $[-50, 50]$ to $[100, 100]$, we could synthesize attack vectors for  PLF, TPF, and TPLF topologies. The effects of the attack vector on those topologies are demonstrated in Fig.~\ref{figPlfVel}-\ref{figTplfPos}. In each of the cases, we can see that the vehicles $\alpha_3$ and $\alpha_4$ face collision due to the FDI attack launched by vehicle $\alpha_2$. From the above result, we observe that a vehicle becomes more resilient towards such FDI attacks as its number of neighbours increases. 
\begin{figure}[!ht]
\centering			\includegraphics[width=\columnwidth,keepaspectratio,clip]{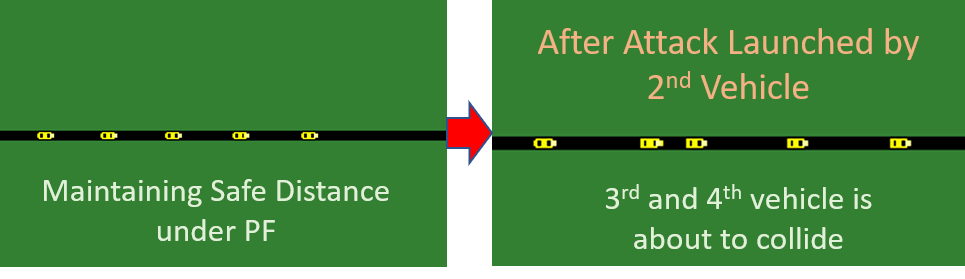}
\vspace{-2em}
\caption{Attack Simulation Using Our Framework}
\label{figAtksim}
\end{figure}
Redundancy of information reduces the effect of the falsified information and hence makes the control strategy more robust. Similar to the safety violating attack vectors, performance violating ones can also be generated using the proposed framework. However, due to space scarcity we only show the $safety$ violating attack simulations. Moreover, in Fig.~\ref{figAtksim}, we can see the output of our simulation module. It demonstrates the safety violating attack scenario on PF topology. We can see that the space between $\alpha_3$ and $\alpha_4$ reduces and leads to a collision as we have seen from the velocity plot with the logged speed data in Fig.~\ref{figPfVel}. We can replicate even more sophisticated falsification scenarios involving more safety-critical maneuvers using the proposed framework. This can be useful to test attack-resilience of the platoon topologies and several state-of-the-art defense mechanisms.

\section{Conclusion}
\label{secConclusion}
In this paper, we present an automated tool-chain that can be used to generate attack vectors for any type of platoon structure and to have a visual simulation of that attack vector on the underlying platoon structure. The tool has been designed on top of an existing connected vehicle simulation tool, namely VENTOS. It gives a provision to customize the dynamics of the vehicles in the platoon. Thus, the proposed tool enables the user not only to test a connected vehicle prototype on varied scenarios but also to design and verify control-theoretic attack detection mechanisms. As future extension of this work, we aim to include customizable attack detection and mitigation modules into the tool-chain.

	% \section*{Acknowledgment}
	
	% The preferred spelling of the word ``acknowledgment'' in America is without 
	% an ``e'' after the ``g''. Avoid the stilted expression ``one of us (R. B. 
	% G.) thanks $\ldots$''. Instead, try ``R. B. G. thanks$\ldots$''. Put sponsor 
	% acknowledgments in the unnumbered footnote on the first page.
	
	\bibliographystyle{IEEEtran}
	\bibliography{reference}
\end{document}